\documentstyle [12pt]{article}
\def\bib{\parskip=0pt\par\noindent\hangindent\parindent
 \parskip =2ex plus .5ex minus .1ex}
\title{The Direct Detection of Non-Baryonic Dark Matter in the Galaxy?}
\author{M.R.S. Hawkins}

\date{}

\begin{document}

\maketitle

{\it
\noindent
Royal Observatory, Blackford Hill, Edinburgh EH9 3HJ, UK\\
}

\begin{abstract}

It has been argued in a number of recent papers that dark matter
is in the form of Jupiter mass primordial black holes which betray
their presence by microlensing quasars.  This lensing accounts for
a number of characteristic properties of quasar light curves, in
both single quasars and gravitationally lensed multiple systems,
which are not explained on the basis of intrinsic variation.  One
prediction of this idea is that Jupiter mass bodies will be detected
by the MACHO experiment as short events of about 2 days duration,
although the expected frequency of detection is still very hard to
estimate.  However, the recent report by the MACHO group of the
detection of a Jupiter mass body in the direction of the Galactic
bulge is consistent with this prediction, and is possibly the first
direct detection of non-baryonic matter in the Galaxy.

\end{abstract}

\section{Introduction}
The idea that non-baryonic dark matter might be in the form of
Jupiter mass primordial black holes has recently been put forward
(Hawkins 1993, 1996) on the basis of an analysis of quasar light
curves.  Long term variations are shown to possess the characteristics
expected from microlensing, such as achromaticity and lack of time
dilation, and a case is made that the variations are not intrinsic
(Hawkins \& Taylor 1997).
A new approach has involved the analysis of multiply lensed quasars,
where microlensing is known to take place (Schild 1996), and the
question is to what extent they constitute typical lines of sight.
It is argued (Hawkins 1997a, 1997b, 1997c) that the stellar population
of the lensing galaxy is not in general capable of producing the
observed variations.\\

If this picture is correct then the mass of the microlensing bodies
can be obtained from the timescale of variation of the quasars and
is found to be $\sim 10^{-3}M_{\odot}$, about the mass of Jupiter.
To produce continuous variation in the quasar light, their mass
density must be around the cosmological critical density.  Baryon
synthesis constraints imply that they must be non-baryonic, and the
most plausible interpretation is that the lenses are primordial black
holes created in the early universe during the quark/hadron phase
transition (Hawkins \& Taylor 1997).\\

\section{MACHO Observations}

On the basis of the quasar observations one can only conclude at
present that the lenses have the critical density on the large
scale.  To probe smaller scales, the MACHO experiment has become
well known as a method for detecting compact bodies in the halo
and bulge of the Galaxy by means of gravitational microlensing
(Alcock et al. 1993).  The idea is to observe the amplification of
distant stars by the rare transits of nearer bodies across the line
of sight.  A detailed description of the technique has recently been
published (Alcock et al. 1997), together with the latest results
in the direction of the Large Magellanic Cloud.  The current conclusion
is that the Galactic halo contains at least 50\% of its mass in the
form of half-solar mass bodies, and the most plausible explanation
seems to be that they are white dwarfs (Adams \& Laughlin 1996), left
over from a very early phase of star formation.\\

Although there is a clear prediction that the Jupiter mass bodies
should eventually be detected in the MACHO survey (Hawkins 1996, 1997d),
the question of the detection frequency has remained an open one as it
has not been clear to what extent non-baryonic dark matter will cluster
into the Galactic halo.  The detection of 50\% of the halo mass in the
form of half-solar mass bodies suggests that it is in fact largely
baryonic, and so the need for non-baryonic matter to cluster on that
scale is lessened, whether it is in the form of SUSY particles or
primordial black holes.\\

As originally conceived, the MACHO programme was most sensitive to
objects in the range $0.1M_{\odot}$ to $0.01M_{\odot}$,
producing events lasting several tens
of days, but the team has recently attempted to extend their coverage
down to planetary mass ($10^{-3}M_{\odot}$) objects giving rise to
microlensing events lasting from 0.3 to 3 days.  This was done by
analysing data from nights where more than one observation of the same
stars was available (Alcock et al. 1996).  As the search for
microlensing events is extended into the planetary mass regime, a
number of new factors come into play.  These include the effect of the
finite size of the source stars, which will eventually decrease the
observed amplitude of the lensing event, and the much smaller number of
samples covering each event.  One would expect a discrete drop in
efficiency as the timescale decreases to two or three days, and most of
the variation takes place in daylight when it cannot be observed.  The
MACHO group claim to have allowed for these effects (Alcock et al.
1996) by estimating their efficiency with a Monte Carlo simulation.
This is certainly a useful approach, but it is not clear how reliable
the results are, as there is no indication of a discrete change in the
probability function as the day and night times become significant.
Nonetheless, in what follows we will adopt the figures published by
the MACHO group.\\

So far, no planetary mass objects have been detected in the halo,
and a limit of around 20\% has been put on their contribution to the
halo mass (Alcock et al. 1997).  However, one such event has
recently been reported in the Galactic bulge where the event rate is
much higher (Bennett et al. 1997).  The event has a duration of two
and a half days, and the MACHO group estimate the mass of the lensing
body to be about 2 $M_{Jup}$, close to the mass found from quasar
microlensing.  The light curve comprises 3 amplified measures in each
colour and appears to be achromatic, but the small number of points
leaves some room for doubting its classification as a microlensing
event.  Again, in what follows we will accept the verdict of the MACHO
team.\\

The event stands well away from the distribution of other (presumably
stellar) lensing events.  The MACHO group estimate that it has only a
small probability of being a part of the tail of stellar mass lenses,
and conclude that it might be a planet in a distant orbit, or one
which has become detached from a planetary system.  We shall see below
that the statistics do not favour this interpretation.
However, the mass and distribution
are as predicted from the analysis of the quasar light curves, where
the population of microlensing bodies is distinct from the stellar
distribution.  If the microlensing picture is correct, then sooner or
later the dark bodies must be detected by the MACHO experiment.  It
thus seems plausible that this short event is the first detection in
the Galaxy of one of the bodies responsible for quasar microlensing
on a larger scale.\\

\section{Discussion}

If the two and a half day MACHO event is caused by one of the Jupiter
mass black holes, it is possible to estimate the density enhancement
over the critical density that this would represent.  First we can
calculate the number of lenses as a fraction of the stellar population.
If we adopt the published detection efficiency as a function of mass
(Alcock et al. 1997) it will be seen that the probability of
detection of a 2 day event is about 0.1 that of a 20 day event.  In
mass terms, the detection of a $0.001M_{\odot}$ body is 10\% as likely
as that of a $0.1M_{\odot}$ body.  To deduce the actual proportion
of $0.001M_{\odot}$ to $0.1M_{\odot}$ bodies one must also allow for
the size of the Einstein disks, which is not incorporated in the
efficiency values since an `event' is defined as a star passing within
the Einstein disk of a lens.  This goes as $\sqrt M$ and thus
effectively decreases the relative detection rate of $0.001M_{\odot}$
bodies by a further factor of 10.  As for the actual detection rate in
the bulge, we see (Bennett et al. 1997) that approximately 100 stellar
mass objects have been detected compared with the one Jupiter
mass body.  Combining these two figures implies that the Jupiter mass
bodies number about the same as the stellar population where the
microlensing is taking place.  If the bodies are detached planets or
planets in distant ($>10 AU$) orbits, the implication is that on
average every star gives rise to such an object.  This seems
implausible and would not appear to be consistent with the
statistics of planetary systems as they are currently understood
(Marcy \& Butler 1997).\\

If we take the stellar space density in the vicinity of the sun at
around 0.12 $pc^{-3}$ (Tinney 1993) and the scale length of the disk as
2.3 kpc (Ruphy et al. 1996), then the space density of stars half way to
the Galactic centre at the most probable position for microlensing is
0.8 $pc^{-3}$.  It has been argued above that the Jupiter mass
objects have the same space density as this, implying a mass
density of about $10^{-3}M_{\odot}\,pc^{-3}$.  Taking the
cosmological critical density as $10^{-7}M_{\odot}pc^{-3}$, this
represents a density enhancement $\delta\rho/\rho \sim 10^{4}$,
comparable to the over-density of galactic halos.
The mass density of the Galactic halo out to 60 $kpc$ is about
$10^{12}M_{\odot}$ (Hawkins 1984), an average mass density of
$10^{-3}M_{\odot}\,pc^{-3}$. This is about the same as the mass
density of Jupiter mass bodies we have deduced from the MACHO
experiment in the direction of the bulge, which if distributed
uniformly could thus account for the dark matter in the halo.
However, a flat halo
profile would be inconsistent with the MACHO results towards the
LMC.  This experiment limits the average space density of
$10^{-3}M_{\odot}$ bodies in the halo to around $10^{-2}pc^{-3}$
(Alcock et al. 1997, Renault et al. 1997),
a factor of 100 down on the density near the bulge.  If the decline
were to follow a power law, then taking into account the integral
constraint of the mass of the halo out to the LMC, it would have an
index of around $-2$, somewhat flatter than the index of $-3$ for halo
stars (Hawkins, 1984).  This is in line with more general expectations
for the distribution of non-baryonic dark matter.  Although the MACHO
results appear to rule out a halo dominated by non-baryonic material
out to around 25kpc, the implication is that it may well predominate
further out.\\

\section{Conclusions}

In this paper we have examined the possibility that the recent
detection of a planetary mass body by the MACHO group could be one
of the Jupiter mass primordial black holes recently proposed as the
constituent of non-baryonic dark matter.  The detection accords with
predictions made on the basis of studies of quasar variability
interpreted as the effects of microlensing.  The mass density
implied by this detection is in line with the density enhancement
associated with galactic halos.\\

\section*{Acknowledgements}

I thank Andy Taylor and Alan Heavens for helpful discussions.  I also
thank the referee Will Sutherland for pointing out the correct
interpretation of some of the MACHO results.\\

\section*{References}

\bib Adams F.C., Laughlin G., 1996, ApJ, 468, 586

\bib Alcock C., {\it et al.}, 1993, Nat, 365, 621

\bib Alcock C., {\it et al.}, 1996, ApJ, 471, 774

\bib Alcock C., {\it et al.}, 1997, ApJ, 486, 697

\bib Bennett D.P., {\it et al.}, 1997, in `Planets Beyond the
 Solar System and the Next Generation of Space Missions', in press

\bib Hawkins M.R.S., 1984, MNRAS, 206, 433

\bib Hawkins M.R.S., 1993, Nat, 366, 242

\bib Hawkins M.R.S., 1996, MNRAS, 278, 787

\bib Hawkins M.R.S., 1997a, in `The Identification of Dark
 Matter', ed. N. Spooner, World Scientific, p. 217

\bib Hawkins M.R.S., 1997b, in `Dark and Visible Matter in
 Galaxies', eds M. Persic \& P. Salucci, ASP Conf. Ser. Vol 117,
 p. 297

\bib Hawkins M.R.S., 1997c, in `18th Texas Symposium on
 Relativistic Astrophysics', in press

\bib Hawkins M.R.S., 1997d, `Hunting Down the Universe',
 Littlebrown, London

\bib Hawkins M.R.S., Taylor A.N., 1997, ApJ, Letters, 482, L5

\bib Marcy G.W., Butler R.P., 1997, in `Brown Dwarfs and
 Extrasolar Planets', in press

\bib Renault C., {\it et al.}, 1997, A\&A, 324, L69

\bib Ruphy S., Robin A.C., Epchtein N., Copet E., Fouqu\'{e} P.,
 Guglielmo F., 1996, A\&A, 313, L21

\bib Schild R.E., 1996, ApJ, 464, 125

\bib Tinney C.G., 1993, ApJ, 414, 279

\end{document}